\documentclass[
	twocolumn,
	twocolappendix,
	]{aastex631}

\usepackage{amsmath}
\usepackage{mathtools} 
\usepackage{dsfont} 

\let\oldtablenum\tablenum 
\let\tablenum\relax
\usepackage{siunitx}
\let\tablenum\oldtablenum

\renewcommand{\d}{\mathrm{d}}
\newcommand{\meanBr}[1]{\left<#1\right>}

\newcommand{\defn}{\equiv}

\newcommand{\abs}[1]{\left|#1\right|}

\newcommand{\im}{\operatorname{Im}}

\renewcommand{\vec}[1]{\boldsymbol{#1}}

\newcommand{\Dirac}{\operatorname{\delta}}
\newcommand{\uvec}[1]{\boldsymbol{\hat{#1}}} 
\newcommand{\pencil}{Pencil} 

\newcommand{\istate}[2]{#1~\textsc{\romannumeral#2}} 
\newcommand{\FT}[1]{\widetilde{#1}} 
\newcommand{\indc}{\operatorname{\mathds{1}}} 

\newcommand{\sgnerr}{\operatorname{sgn}_{\Delta}}
\newcommand{\err}{\Delta} 
\newcommand{\kmax}{k_\text{max}} 

\newcommand{\Hspec}{\widetilde{H}} 
\newcommand{\Espec}{\widetilde{E}} 

\newcommand{\BLOS}{B_\textsc{LOS}} 

\newcommand{\Textcite}{\Citet}
\newcommand{\parencite}{\citep}

\usepackage[final]{listings}
\usepackage{color}
\definecolor{mygreen}{RGB}{28,172,0} 
\definecolor{mylilas}{RGB}{170,55,241}
\usepackage[T1]{fontenc} 
\usepackage{textcomp} 
\usepackage{accsupp} 

\newlength{\lstcolumnwidth}
\newcommand\emptyaccsupp[1]{\BeginAccSupp{ActualText={}}#1\EndAccSupp{}}
\lstdefinestyle{python}{language=Python,
	upquote=true,
	basicstyle={\ttfamily},
	basewidth=\lstcolumnwidth,
	columns=fixed,
	fontadjust=true,
	morekeywords={matlab2tikz},
	keywordstyle=\color{blue},
	morekeywords=[2]{1}, keywordstyle=[2]{\color{black}},
	identifierstyle=\color{black},
	stringstyle=\color{mylilas},
	commentstyle=\color{mygreen},
	showstringspaces=false,
	breaklines=true,
	numbers=left,
	numberstyle={\tiny \color{black} \emptyaccsupp},
	numbersep=9pt, 
	}

\lstdefinestyle{default}{
	upquote=true,
	basicstyle={\ttfamily},
	identifierstyle=\color{black},
	stringstyle=\color{mylilas},
	commentstyle=\color{mygreen},
	showstringspaces=false,
	breaklines=true,
	numbers=left,
	numberstyle={\tiny \color{black} \emptyaccsupp},
	numbersep=9pt, 
	}
\shorttitle{Solar magnetic energy and helicity spectra}
\shortauthors{Gopalakrishnan et al.}

\begin{document}

\title{The spectra of solar magnetic energy and helicity}

\correspondingauthor{G. Kishore}

\author[0000-0003-2620-790X]{G. Kishore}
\email{kishoreg@iucaa.in}

\author[0000-0001-6097-688X]{Nishant K. Singh}
\email{nishant@iucaa.in}
\affiliation{Inter-University Centre for Astronomy \& Astrophysics, Post Bag 4, Ganeshkhind, Pune 411 007, India}

\begin{abstract}
	Previous studies have used magnetic energy and helicity spectra, the latter computed using the two-scale method, to search for signatures of mean-field dynamos.
	In this study, we compare cotemporal HMI and SOLIS synoptic vector magnetograms to illustrate the instrument-dependence of even qualitative features of the energy and helicity spectra.
	Around the minimum between solar cycles 24 and 25, we find that the magnetic energy spectrum computed from HMI observations exhibits two distinct peaks.
	One of these peaks is only present near the cycle minimum, and corresponds to large-scale magnetic fields at high latitudes.
	Nevertheless, such magnetic fields are not present in contemporaneous synoptic vector magnetograms from SOLIS.
	Further, even when magnetograms from both the instruments are apodized, the helicity spectra calculated using the two-scale method disagree (on both the sign and the value of the fractional helicity).
	This suggests that currently available synoptic magnetograms are not reliable enough for such studies.
\end{abstract}


\section{Introduction}

The Sun exhibits a magnetic cycle that is roughly 22 years long, along with a correlated 11-year sunspot cycle \citep{Hat15}.
While the factors determining the strength and the periodicity of the solar magnetic field are a matter of debate, most proposed explanations appeal to dynamos of some kind or the other \citep{charbonneau2020}.

Magnetic helicity (a measure of the topological linkage between magnetic field lines) is conserved in ideal magnetohydrodynamics; further, the rate of its dissipation approaches zero in the astrophysically relevant limit of large magnetic Reynolds number (\citealp[section 3.4]{KanduPhysicsReports2005}; \citealp{ParValDem15}).
It thus plays a crucial role in mean-field dynamo theory.

The kinetic helicity in the Sun is theoretically expected to be negative in the northern hemisphere and to flip sign across the equator \citep[p.~156]{RobSti71}.
For a large-scale magnetic field generated exclusively through the $\alpha$ effect, one expects the sign of the associated magnetic helicity to be the opposite of that of the kinetic helicity.
The resulting change in the sign of the large-scale magnetic helicity across the equator is referred to as the \emph{hemispheric sign rule} (HSR).
Further, in such a scenario, the small- and large-scale fields in a particular hemisphere are expected to have magnetic helicities of opposite signs, such that the magnetic helicity of the total magnetic field in a particular hemisphere is close to zero \citep[see][eqs.~9.11, 9.12]{KanduPhysicsReports2005}.
On the other hand, in a mean-field model that also includes differential rotation, \citet{PipZhaSok13} found reversals of the HSR near cycle minima.
The spatial and spectral distribution of the magnetic helicity is thus a source of information on which dynamo mechanisms contribute to the solar magnetic cycle.

One way to observationally test the HSR is to use the helicity of the electric current in active regions as a proxy for the magnetic helicity therein.
Applying this method to observations of a number of active regions, \citet{See90} and \citet{ZhaSakPev10} found latitudinal sign changes corresponding to those expected for the small-scale magnetic helicity.

Another approach, first used by \citet{BraBlaSar03}, is to calculate the magnetic helicity of the axisymmetric part of the solar magnetic field from the longitudinal average of observed synoptic magnetograms.
Using a similar technique, \citet{PipPev14} found that the helicity of the axisymmetric part of the magnetic field obeys the HSR corresponding to large-scale fields.
They also found that while the magnetic and current helicities of the large-scale fields show broadly similar latitudinal trends, they differ in detail.

\Citet{PipPevLiu19}, who computed the helicity densities of both the axisymmetric and non-axisymmetric parts of the magnetic field, found that while the axisymmetric part behaves as expected for a large-scale field, the non-axisymmetric part does not consistently display the behaviour expected from the HSR for a small-scale field (i.e., the helicities of the axisymmetric and non-axisymmetric parts are occasionally of the same sign).

While the studies mentioned above considered any non-axisymmetric part of the magnetic field as a small-scale component, a more nuanced approach based on the helicity spectrum was used by \citet{ZhaBraSok14}.
Computing the magnetic helicity spectrum over a single active region, they found a sign consistent with the HSR at large scales.
Applying the same method to thousands of active regions, \citet{GosBra19} found that individual active regions only obey the HSR poorly.

A complication in calculating global helicity spectra (as opposed to spectra over isolated active regions) is that the expected sign change of the helicity across the equator would lead to the helicity spectrum being zero at each wavenumber if calculated over the entirety of the Sun's surface.
To capture the expected sign change across the equator, \citet{BraPetSin17} applied a `two-scale' method to synoptic magnetograms in order to calculate energy and helicity spectra.
They used a Cartesian approximation to facilitate the application of Fourier transforms.
\citet{PraSinKap21} have described a similar method that uses spherical harmonics, and shown that the results are similar to those obtained in the Cartesian approximation.
We will thus continue to use the Cartesian approximation in this work.

Comparing synoptic magnetograms from different facilities (including HMI,\footnote{
Helioseismic and Magnetic Imager;
\url{http://hmi.stanford.edu} 
} MDI,\footnote{
Michelson Doppler Imager;
\url{http://soi.stanford.edu} 
} and SOLIS\footnote{
Synoptic Optical Long-term Investigations of the Sun;
\url{https://nso.edu/telescopes/nisp/solis} 
}), \citet{RilBenLin14} found that conversion factors between the line-of-sight magnetic fields measured by different instruments can be latitude-dependent, and even time-dependent.
Such differences are expected to affect the energy and helicity spectra.
Further, comparing synoptic magnetograms from HMI and MDI, \citet{LuoJiaWan23} have found that the energy spectra can only be matched by applying a scale-dependent correction factor.
Due the theoretical importance of the sign of the helicity spectrum, one would like to understand how strongly it is affected by such instrument-dependent effects.

Dividing full-disk vector magnetograms from the Huairou Solar Observing Station
into strong-field ($\abs{\BLOS} > \qty{1000}{G}$, where $\BLOS$ is the line-of-sight component of the magnetic field) and weak-field ($\qty{100}{G} < \abs{\BLOS} < \qty{500}{G}$) parts and analysing them separately, \citet{Zha06} found that their current helicities are of opposite signs, with the weak fields obeying (only in a statistical sense) the HSR corresponding to small-scale fields.
On the other hand, \textcite{GosPevRud13}, who performed a similar study using SOLIS synoptic vector magnetograms, arrived at the conclusion that it is the strong fields that obey the HSR corresponding to small-scale fields, with the weak field regions having current helicities of the opposite sign.
One possible explanation for this disagreement is that one or both of these observations are affected by instrument-dependent systematic biases.
Another possible explanation is that biases introduced during the assembly of synoptic magnetograms affect inferences of the magnetic helicity and its proxies.

Using the two-scale method in the Cartesian approximation, \citet{SinKapBra18} have found that during Carrington rotations 2161--2163, the sign of the helicity spectrum as computed from SOLIS synoptic magnetograms disagrees with that computed from HMI synoptic magnetograms.
However, it is unclear whether this particular interval of time is an aberration, or if helicity spectra computed from synoptic magnetograms using the two-scale method are always instrument-dependent.

Our main goal in this study is to compare the helicity spectra computed from HMI and SOLIS synoptic vector magnetograms in Carrington rotations 2097--2195 (slightly more than 7 years), and check how well they agree.
In the process, we clarify various implementation details and correct errors in previous applications of the two-scale method.
We also show that around the solar minimum, discrepancies in the polar fields are strong enough to qualitatively change even the magnetic energy spectrum.

Section \ref{bihelical: section: describe two-scale} describes the two-scale method.
Since previous implementations of the two-scale method had various errors, we verify our implementation by applying it to simulations in section \ref{bihelical: section: verify two-scale}.
Section \ref{bihelical: section: data analysis} gives the details of the data we have used and our analysis procedure.
In section \ref{bihelical: section: HMI large-scale fields}, we show that near the minimum between solar cycles 24 and 25, one can find large-scale magnetic fields at high latitudes in HMI synoptic magnetograms, but not in cotemporal SOLIS synoptic magnetograms.
Section \ref{bihelical: section: HMI SOLIS helspec} shows that even when the high latitudes are removed, the helicity spectra calculated from HMI and SOLIS synoptic magnetograms do not agree.
We show that agreement is bad throughout the time interval in which data are available from both the instruments.
Finally, section \ref{bihelical: section: conclusions} summarizes our conclusions.
The scripts used for this study, along with the input files for the simulations, have been archived on Zenodo \citep{zenodo_bihelical_2025}; alternatively, they can be downloaded from \url{https://github.com/Kishore96in/bihelical-mag-fields}.

\section{Description of the two-scale method}
\label{bihelical: section: describe two-scale}

\subsection{Locally isotropic, weakly inhomogeneous turbulence}

We are interested in the two-point correlation function of the magnetic field: $\meanBr{B_i(\vec{x}) \, B_j(\vec{y}) }$.
Rather than using the spatial variables $\vec{x}$ and $\vec{y}$ directly, it is convenient to use $\vec{r} \defn \vec{x} - \vec{y}$ and $\vec{R} \defn ( \vec{x} + \vec{y} )/2$;
this is because in homogeneous turbulence, the two-point correlation function is independent of $\vec{R}$.
Once the two-point correlation function has been written in terms of $\vec{r}$ and $\vec{R}$, its Fourier transform\footnote{
In this work, we use the convention that in $d$ spatial dimensions, the forward Fourier transform contains $(2\pi)^{-d}$.
} can be expressed as
\begin{equation}
	M_{ij}(\vec{k}, \vec{K})
	\defn
	\meanBr{\FT{B}_i{\left(\vec{k} + \tfrac{1}{2} \, \vec{K} \right)} \, \FT{B}_j^*{\left(\vec{k} - \tfrac{1}{2} \, \vec{K} \right)} }
	\label{bihelical: eq: Mij defn}
\end{equation}
where $\vec{k}$ is the Fourier conjugate of $\vec{r}$, and $\vec{K}$ is the Fourier conjugate of $\vec{R}$.

For turbulence that is locally isotropic and weakly inhomogeneous,
$M_{ij}$ is completely specified by the functions \citep[see][eqs.~2.19 and 2.23]{robertsSoward75},
\begin{align}
	\begin{split}
		8 \pi M(k,\vec{K})
		={}&
		\int M_{ii}(\vec{k}, \vec{K}) \, \d\Omega_k
		\label{bihelical: eq: mag spec 3D}
	\end{split}
	\\
	\begin{split}
		8\pi N(k, \vec{K})
		={}&
		\int i k_i \epsilon_{ijk} \, M_{jk}(\vec{k}, \vec{K}) \, \d\Omega_k
		\label{bihelical: eq: maghel spec 3D}
	\end{split}
\end{align}
These are related to the quantities $\widetilde{E}_M$ and $\widetilde{H}_M$ (defined by \citealp{ZhaBraSok14}; \citealp{BraPetSin17}; and \citealp{SinKapBra18}) by
\begin{subequations}
\begin{align}
	\begin{split}
		\widetilde{E}_{M,\text{3D}}(k, \vec{K})
		={}&
		4\pi k^2 \, M(k, \vec{K})
	\end{split}
	\\
	\begin{split}
		={}&
		\frac{1}{2} \int M_{ii}(\vec{k}, \vec{K}) \, k^2 \, \d\Omega_k
	\end{split}
	\\
	\begin{split}
		\widetilde{H}_{M,\text{3D}}(k, \vec{K})
		={}&
		8\pi\, N(k, \vec{K})
	\end{split}
	\\
	\begin{split}
		={}&
		\int i \, \epsilon_{ijk} \, \frac{k_i}{k^2} \, M_{jk}(\vec{k}, \vec{K}) \, k^2 \, \d\Omega_k
	\end{split}
\end{align} \label{bihelical: eq: EM HM defn 3D}%
\end{subequations}
Note that the magnetic energy and the magnetic helicity are proportional to $\int\widetilde{E}_{M,\text{3D}} \ \d k$ and $\int\widetilde{H}_{M,\text{3D}} \, \d k$ respectively.

\subsection{Definitions of the energy and helicity spectra}

While our reasoning so far has been in terms of the continuous Fourier transform, it is conventional in observational work to use the discrete Fourier transform.
If we define $M_{ij}$ (equation \ref{bihelical: eq: Mij defn}) in terms of the discrete Fourier transform of the magnetic field, we find that it has the same dimensions as the square of the magnetic field.
Analogous to equations \ref{bihelical: eq: EM HM defn 3D}, we define the energy and helicity spectra as
\begin{align}
	\begin{split}
		\Espec(k, \vec{K})
		\defn{}&
		\sum_{\vec{q}} \frac{1}{2} \, M_{ii}(\vec{q}, \vec{K}) \indc(\vec{q}, k)
	\end{split} \label{B3.bihelical: eq: EM definition}
	\\
	\begin{split}
		\Hspec(k, \vec{K})
		\defn{}&
		\sum_{\vec{q}} i \, \epsilon_{ijk} \, \frac{q_i}{q^2} \, M_{jk}(\vec{q}, \vec{K}) \indc(\vec{q}, k)
	\end{split} \label{B3.bihelical: eq: HM definition}
\end{align}
where $\indc(\vec{q}, k)$ is one when $- \Delta/2 \le \abs{\vec{q}} - k < \Delta/2$, and zero elsewhere ($\Delta$ is the spacing between adjacent values of $k$); note that this indicator function has replaced the 3D angular element $k^2 \, \d\Omega_k$.
Note that $\Espec$ and $k \Hspec$ have units of \si{G^2}, which is equivalent to \si{erg.cm^{-3}}.\footnote{
\textcite{BraPetSin17}, \textcite{SinKapBra18}, and \textcite{PraSinKap21} mention dimensionally different units (\si{G^2.Mm}).
Their expressions would follow if the spectra were defined using the continuous Fourier transform, as opposed to the discrete one used here.
We note that at least \textcite{BraPetSin17} used the discrete Fourier transform in their scripts.
}

\subsection{Extension to 2D}
In principle, the expressions above are not directly applicable to solar magnetograms, since they require information about the magnetic field in a three-dimensional volume.\footnote{
This problem is not unique to the two-scale method; it affects all the methods described in the introduction.
}
\Citet[p.~3]{ZhaBraSok14} and \citet[p.~2]{BraPetSin17} heuristically derived their expressions by making the replacement $4\pi k^2 \, \d\Omega_\text{3D} \to 2\pi k \, \d\Omega_\text{2D}$ in equations \ref{bihelical: eq: EM HM defn 3D}.
Similarly, we use equations \ref{B3.bihelical: eq: EM definition} and \ref{B3.bihelical: eq: HM definition}, but with the sum over $\vec{q}$ restricted such that the component of $\vec{q}$ out of the plane is zero (i.e.\@ we only sum over wavevectors lying within the plane).

\subsection{The important part of the helicity spectrum}
\label{section: negative imaginary part}

According to the HSR, the large-scale magnetic fields have helicities of opposite signs in the northern and southern hemispheres.
If one is searching for signs of large-scale dynamo action, one should thus look at the part of the large-scale helicity that changes sign about the equator.
An example of such a function is $\sin(2\lambda)$, with $\lambda$ being the latitude in radians.
Recalling the Fourier transform
\begin{equation}
	\sin( \vec{K}_1 \cdot \vec{X}) \to \frac{1}{2i} \left[ \Dirac{\!\left(\vec{K} - \vec{K}_1 \right)} - \Dirac{\!\left(\vec{K} + \vec{K}_1 \right)} \right]
\end{equation}
we find that the part of the helicity spectrum that corresponds to such a modulation has a large-scale wavevector $\vec{K}_1 \defn 2\pi \uvec{\lambda} /L_\lambda$.
Further, to obtain the contribution proportional to this with a sign corresponding to the sign of the helicity in the northern hemisphere, one has to take the \emph{negative imaginary part} of $\Hspec(k, \vec{K}_1)$.

\section{Verification of the two-scale method}
\label{bihelical: section: verify two-scale}

\Citet{BraPetSin17} attempted to verify their implementation of the two-scale method by applying it to slices saved from a simulation.
Examination of the script they used to create their figure 6b\footnote{
\url{https://lcd-www.colorado.edu/~axbr9098/projects/LShelicityspec/576b/ptst.pro} (accessed 26-Aug-2023, 11:16 AM IST).
} suggests that what they plot is actually the \emph{positive} imaginary part of the helicity spectrum (and not the relevant negative imaginary part as claimed).
We now check if the two-scale method remains valid once this (and another mistake which we will point out in due course) is corrected.

Our simulation setup is similar to that used by \textcite[sec.~3.2]{BraPetSin17}.
We consider isothermal turbulence in a periodic cube of size $2\pi$ length units in each direction.
The velocity field is forced by linear combinations of the eigenfunctions of curl operator, chosen to have a prescribed fractional helicity \citep[see][]{Bra01, HauBraMee04}, with the helicity switching sign across $z=0$ (the midplane) if necessary.
The wavevector of the forcing function is randomly chosen at each timestep to be in the shell $9.5 \le k < 10.5$.
The evolution equations are solved using the \pencil{} code \parencite{Pencil2021}.\footnote{
\url{https://pencil-code.nordita.org}
}
The spatial discretization uses $512^3$ points and a sixth-order finite difference scheme.
The equations are evolved in time using a third-order Runge-Kutta method.
Upwinding is used for advection of the density, the velocity, and the magnetic vector potential \parencite[appendix B]{DobStiBra06}.

Given the magnetic field at a particular instant of time, we consider its values in a single $yz$ plane, and calculate the energy and helicity spectra.\footnote{
Taking $xz$ planes, like \textcite[sec.~3.2]{BraPetSin17}, forces one to use a left-handed coordinate system (this can be compensated for as described in appendix \ref{B3.bihelical: section: helspec_two mistake}).
}
This coordinate system is similar to the one we will later use for synoptic maps (section \ref{B3.bihelical: section: pseudo-cartesian coordinates}).

\subsection{No sign change of helicity}
\begin{figure}
	\centering
	\includegraphics{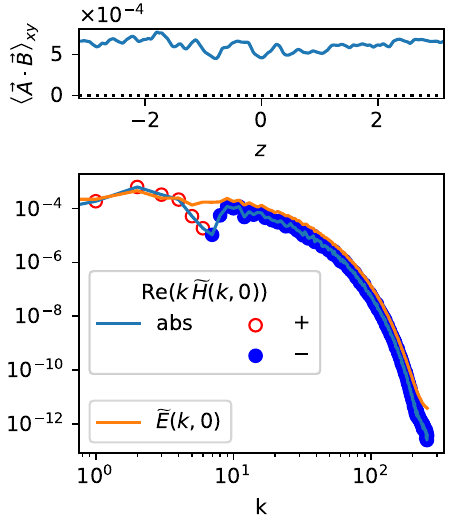}
	\caption{
	Recovery of the correct sign of the helicity from a simulation where there is no sign change across $z=0$.
	The helicity in the top panel, calculated directly from $\vec{A}$ and $\vec{B}$, has been averaged over $x$ and $y$.
	The helicity spectrum in the bottom panel is the average of the spectra calculated from $\vec{B}$ in six $yz$ planes.
	The solid blue line indicates the absolute value of the relevant part of the helicity spectrum, while an open red (filled blue) circle denotes a positive (negative) sign at that wavenumber.
	None of the quantities has been averaged over time.
	We see that the sign of the helicity in the top panel matches the sign of the helicity spectrum at low wavenumbers in the bottom panel.
	}
	\label{B3.bihelical.validation: fig: sim 2}
\end{figure}
First, we consider a case where the sign of the helical forcing is the same throughout the domain.
In this case, we are interested in the real part of the $K=0$ part of the helicity spectrum.
Figure \ref{B3.bihelical.validation: fig: sim 2} shows that we recover the correct sign of the helicity spectrum.

\subsection{Helicity changing sign across the equator}
Next, we consider a case where the helicity flips sign across $z=0$, and the domain itself has extent $-L_z/2 < z < L_z/2$.
In this case, we expect the strongest contribution to be proportional to $\sin(2 \pi z/ L_z)$ (see section \ref{section: negative imaginary part}).

\subsubsection{The problem}
Formally, the expressions for the energy and helicity spectra (equations \ref{B3.bihelical: eq: EM definition} and \ref{B3.bihelical: eq: HM definition}) involve $ M_{ij}(\vec{k}, \vec{K}) \defn \meanBr{\FT{B}_i{\left(\vec{k} + \tfrac{1}{2} \, \vec{K} \right)} \, \FT{B}_j^*{\left(\vec{k} - \tfrac{1}{2} \, \vec{K} \right)} } $.
If one wishes to recover a mode with $K_z = 2\pi/L_z$, a problem arises: $2\pi/L_z$ is the smallest wavenumber present in the domain, and so it is difficult to make sense of $\vec{K}/2$.

\begin{figure}
	\centering
	\includegraphics{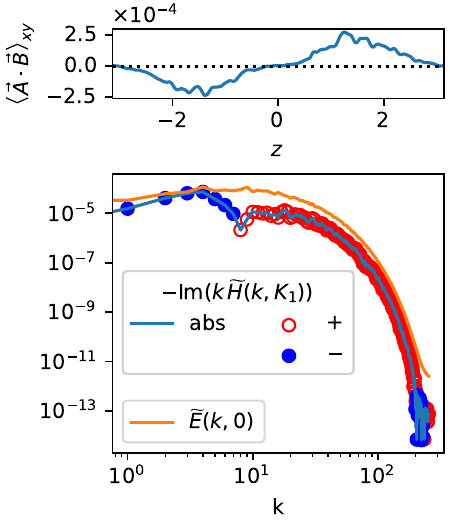}
	\caption{
	Similar to figure \ref{B3.bihelical.validation: fig: sim 2}, but using a simulation where the helicity changes sign across $z=0$.
	The sign of the negative imaginary part of the magnetic helicity spectrum at low wavenumbers (lower panel) is expected to match the sign of the mean magnetic helicity at $z>0$ (upper panel).
	However, we find that the calculated magnetic helicity spectrum has the wrong sign when one takes $M_{ij}(\vec{k}, \vec{K}) \defn \meanBr{\FT{B}_i{\left(\vec{k} + \vec{K} \right)} \, \FT{B}_j^*{\left(\vec{k} \right)} } $ as in previous studies.
	}
	\label{B3.bihelical.validation: fig: sim 1 one-side shift wrong sign}
\end{figure}

\subsubsection{The wrong method}
\label{B3.bihelical.valication: section: traditional}
To circumvent this issue, \textcite{BraPetSin17} and \textcite{SinKapBra18} treat the quantity $ \meanBr{\FT{B}_i{\left(\vec{k} + \vec{K} \right)} \, \FT{B}_j^*{\left(\vec{k} \right)} } $ as being equivalent to $M_{ij}(\vec{k}, \vec{K})$.
If one uses this definition, another mistake (appendix \ref{B3.bihelical: section: helspec_two mistake}) needs to be made to recover the correct sign of the helicity spectrum.
In figure \ref{B3.bihelical.validation: fig: sim 1 one-side shift wrong sign}, we show that the wrong sign of the helicity is recovered using this method.

\begin{figure}
	\centering
	\includegraphics{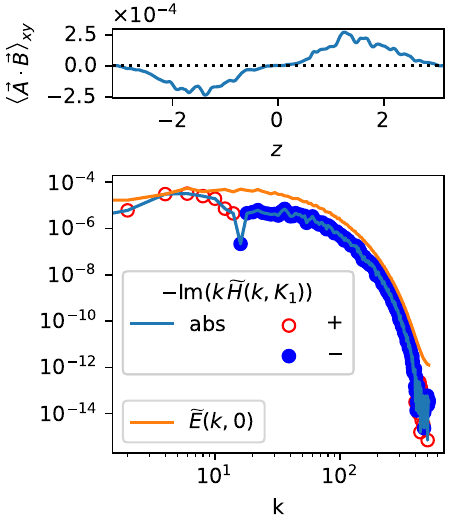}
	\caption{
	Analysis of the same simulation as in figure \ref{B3.bihelical.validation: fig: sim 1 one-side shift wrong sign}, but now taking $M_{ij}(\vec{k}, \vec{K}) \defn \meanBr{\FT{B}_i{\left(\vec{k} + \vec{K}/2 \right)} \, \FT{B}_j^*{\left(\vec{k} - \vec{K}/2\right)} } $.
	The magnetic helicity spectrum now has the correct sign.
	}
	\label{B3.bihelical.validation: fig: sim 1 doubledomain correct sign}
\end{figure}

\subsubsection{The correct method}
An alternative
is to double the length of the domain (keeping the equator, if any, still at the middle); this is consistent with the assumption of periodicity that goes into expanding a function (in a finite domain) as a Fourier series.
In this particular case, we extend the definition of $\vec{B}$ as (omitting the other arguments for brevity)
\begin{equation}
	\vec{B}(z) =
	\begin{dcases}
		\vec{B}(z + L_z) \,, & -L_z \phantom{/2} < z < -L_z/2 \\
		\vec{B}(z) \,, & -L_z/2 < z < \phantom{-} L_z/2 \\
		\vec{B}(z - L_z) \,, & \phantom{-} L_z/2 < z < \phantom{-} L_z
	\end{dcases}
	\label{B3.bihelical.validation: eq: doubled domain definition}
\end{equation}
To keep the spacing between the wavevectors the same in both the directions, we also double the domain along the $y$ direction (recall that the domain was initially cubical).\footnote{
In the $y$ direction, we can simply stack the domain, since there is no equator to worry about.
}
After calculating the spectra, we rebin them in terms of the wavenumbers of the original domain (since doubling the domain halves the spacing between the wavenumbers).
In figure \ref{B3.bihelical.validation: fig: sim 1 doubledomain correct sign}, we show that this method allows us to recover the correct sign of the helicity spectrum.

\section{Data analysis}
\label{bihelical: section: data analysis}

\subsection{The magnetograms used}
\subsubsection{HMI}
\label{bihelical: section: HMI magnetograms}

HMI is an instrument on SDO (Solar Dynamics Observatory, a geosynchronous satellite) which observes the \SI{617.3}{nm} line of \istate{Fe}{1} in absorption.\footnote{
For details on the properties of this particular line, we direct the reader to \textcite{NorGraUlr06}.
}
After computing the Stokes parameters from these measurements, the vector magnetic field on the observed solar disk is obtained using a Milne-Eddington-based inversion algorithm.
Azimuth disambiguation is done using the minimum-energy algorithm in strong-field ($\abs{\vec{B}} > \qty{150}{G}$) regions, while the sign of the transverse component is randomly assigned in weak-field regions.
Synoptic maps are then constructed by, for each combination of Carrington longitude and latitude, averaging over magnetograms where that particular location is closest to the observed central meridian.
\Textcite{LiuHoeSun17} describe the production of both full-disk and synoptic vector magnetograms in more detail.

We use $360\times720$ pixel (sine-latitude vs longitude) vector synoptic maps (series \lstinline{b_synoptic_small}) downloaded from JSOC.\footnote{
\url{http://jsoc.stanford.edu/} (accessed on the 1st of April, 2023)
}
We rebin these using cubic spline interpolation to a grid equispaced in latitude, preserving the number of pixels.

\subsubsection{SOLIS}
SOLIS's VSM (Vector Spectro-Magnetograph) observes a pair of \istate{Fe}{1} lines around \qty{630.2}{nm} in absorption \parencite[section 2]{KelHarGia03}.
The vector magnetic field is obtained from the Stokes parameters using a Milne-Eddington-based inversion algorithm \parencite{Har17}.
Azimuthal disambiguation has supposedly been done using the `Very Fast Disambiguation method' proposed by \textcite{RudAnf14}.\footnote{
We have not been able to confirm this through primary sources, and so rely on secondary sources such as \textcite[section 3]{SinKapBra18}.
}
Synoptic maps are constructed by averaging over multiple magnetograms, with a weighting factor ensuring that the dominant contributions to a particular Carrington longitude come from magnetograms where that longitude is near the central meridian.
We use $180\times360$ pixel (latitude vs longitude) vector synoptic maps.\footnote{
Downloaded from \url{https://magmap.nso.edu/solis} (series \lstinline{kcv9g}) on the 8th of September, 2023.
These files are described as ``level 3 SOLIS data processed by GONG pipeline''.
\textcite{GosPevRud13} appear to describe the creation of these synoptic maps in more detail, but also see
\url{https://solarnews.nso.edu/solis-vsm-vector-magnetograms} 
and
\url{https://solis.nso.edu/0/vsm/aboutmaps.html}. 
}

\subsection{Coordinate system and domain}
\label{B3.bihelical: section: pseudo-cartesian coordinates}

Using $\uvec{r}$, $\uvec{\theta}$, and $\uvec{\phi}$ to denote the radial, colatitudinal, and azimuthal unit vectors on the Sun's surface, we note that $\uvec{r}$, $\uvec{\phi}$, and $\uvec{\theta}$ form a left-handed coordinate system.
A right-handed coordinate system can then be formed by taking one's unit vectors as $\uvec{r}$, $\uvec{\phi}$, and $\uvec{\lambda}$, where $\uvec{\lambda} \defn -\uvec{\theta}$.
The components of the magnetic field are given by $(B_r, B_\phi, B_\lambda) = (B_r, B_\phi, - B_\theta)$ (\citealp[eq.~20]{BraPetSin17}; \citealp[eq.~10]{SinKapBra18}).

We unwrap the surface of the Sun to form a Cartesian 2D domain, such that the position vector is $\vec{X} = (X_\phi, X_\lambda)$; and the domain has size $L_\phi \times L_\lambda = 2\pi R_\sun \times \pi R_\sun$ (where $\lambda \defn \pi/2 - \theta$ is the latitude).
As discussed in section \ref{section: negative imaginary part}, the large-scale wavevector we are interested in is then $\vec{K}_1 \defn 2\pi \uvec{\lambda} /L_\lambda$.

\subsection{Averaging}

Recall that the two-point correlation of the magnetic field (equation \ref{bihelical: eq: Mij defn}) is defined as an average.
In the context of this work, the magnetic field used to calculate the two-point correlations is from synoptic vector magnetograms, and the averaging is over the azimuthal large-scale coordinate (due to our choice of the large-scale wavevector), the angular part of the small-scale coordinate, and time.

The magnetic field in a synoptic magnetogram may itself be thought of as an average over multiple full-disk observations, meaning that some information on the two-point correlations is lost in the process of creating synoptic magnetograms.
Exploring how this affects the calculated spectra is left to future work.

\subsection{Error estimates}
Spectra calculated from single magnetograms are expected to be affected by stochastic fluctuations.
To suppress these fluctuations, we average spectra over a number of contiguous magnetograms.
We then estimate the stochastic errors in these averaged spectra by using the jackknife method: given $N$ realizations of a random variable $X$ (denoted by $X_1, \dots X_N$), the error in their mean is estimated as the standard deviation of $\mu_1, \dots \mu_N$, where
\begin{equation}
	\mu_i \defn \sum_{\substack{j=1\\j \ne i}}^N \frac{X_j}{N-1}
	\label{bihelical: eq: error estimate}
\end{equation}

The procedure chosen for azimuth disambiguation is expected to affect the helicity spectrum.
In particular, for the HMI magnetograms, the sign of the transverse field is randomly chosen in weak-field regions.
Since the random signs chosen for different magnetograms are uncorrelated, the error estimated using equation \ref{bihelical: eq: error estimate} also includes the error due to this specific disambiguation procedure.
We show this in appendix \ref{bihelical: appendix: azimuth error}.
However, note that our error estimate does not include the error due to other disambiguation methods (e.g.\@ the minimum energy algorithm used for the HMI magnetograms in strong-field regions).

No attempt is made to estimate other sources of systematic error, e.g.\@ due to the model atmospheres used in the inversion process.

\section{Signs of polar fields in magnetic energy spectra}
\label{bihelical: section: HMI large-scale fields}

\begin{figure}
	\centering
	\includegraphics{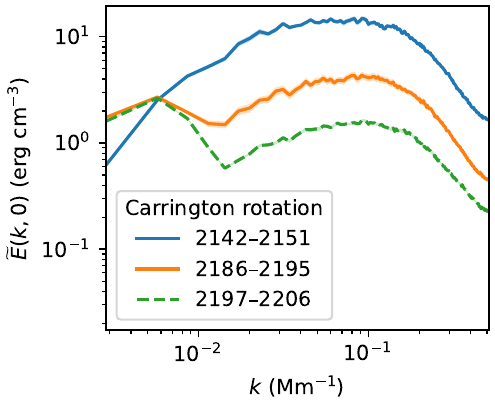}
	\caption{
		Energy spectra calculated from HMI magnetograms, averaged over the indicated Carrington rotations.
		Errors are indicated by shaded regions of the same colour as the corresponding lines, but are too small to be clearly visible.
		}
	\label{bihelical: fig: HMI energy spectra twopeak}
\end{figure}
\begin{figure}
	\centering
	\includegraphics{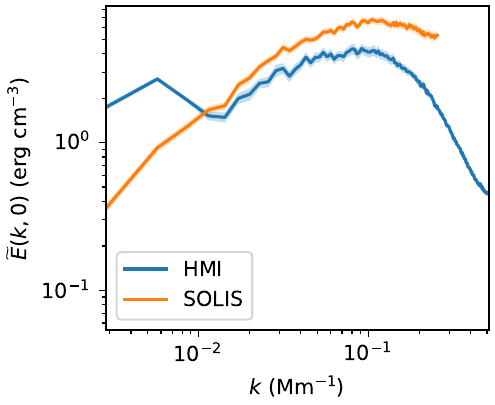}
	\caption{
		Energy spectra calculated from HMI and SOLIS magnetograms, averaged over Carrington rotations 2186--2195.
		In the case of SOLIS alone, Carrington rotation 2192 has been excluded due to poor data coverage.
		Errors are indicated by shaded regions of the same colour as the corresponding lines.
		}
	\label{bihelical: fig: HMI energy spectra twopeak vs SOLIS}
\end{figure}
\begin{figure}
	\centering
	\includegraphics{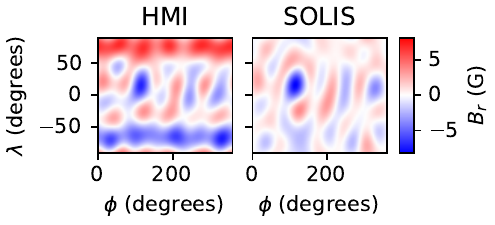}
	\caption{
		Low-pass-filtered ($k < \qty{e-2}{Mm^{-1}}$) synoptic HMI and SOLIS maps of the radial magnetic field in Carrington rotation 2195 (12-Sep-2017 -- 09-Oct-2017).
		}
	\label{bihelical: fig: HMI synoptic lowpass}
\end{figure}
\begin{figure}
	\centering
	\includegraphics{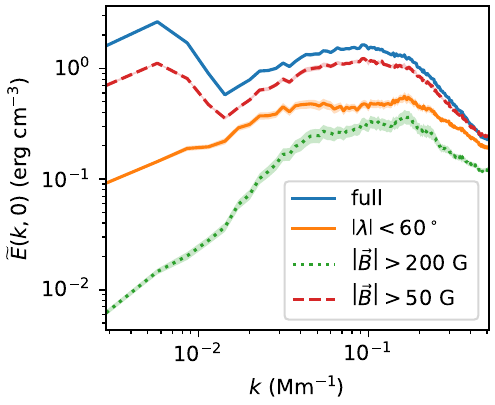}
	\caption{
		The effect, of removing high latitudes or weak-field regions from HMI magnetograms, on the magnetic energy spectrum (averaged over Carrington rotations 2197--2206).
		The legend indicates which region of each magnetogram was used to compute the spectrum.
		Errors are indicated by shaded regions of the same colour as the corresponding lines.
		}
	\label{bihelical: fig: HMI effect apodization masking}
\end{figure}

Figure \ref{bihelical: fig: HMI energy spectra twopeak} shows the magnetic energy spectra calculated from HMI synoptic vector magnetograms at different phases of solar cycle 24.
Near the peak of solar cycle 24
(Carrington rotation 2149),
we find that the magnetic energy spectrum peaks at a single scale.
However, closer to the minimum between cycles 24--25
(Carrington rotation 2224),
we find that the magnetic energy spectrum has a second peak at around \qty{6e-3}{Mm^{-1}}.
Figure \ref{bihelical: fig: HMI energy spectra twopeak vs SOLIS} shows that such a peak is not present in energy spectra calculated from SOLIS synoptic vector magnetograms.

Let us now investigate what exactly these large-scale features correspond to.
The left panel of figure \ref{bihelical: fig: HMI synoptic lowpass} shows a low-pass-filtered HMI synoptic map from a single Carrington rotation (2195).
The large-scale features we saw in the spectra show up as a dipolar component, strongest at latitudes $\abs{\lambda} > \ang{50}$.
The sign of this large-scale dipolar field is consistent with that expected for the polar fields at this stage of the solar cycle \parencite[see][figure 17]{Hat15}.

The right panel of figure \ref{bihelical: fig: HMI synoptic lowpass} shows a low-pass-filtered SOLIS magnetogram.
While the magnetograms from HMI and SOLIS are comparable near the equator, they are completely different at high latitudes.
In particular, the SOLIS magnetograms do not have a prominent dipolar component at high latitudes.
Given the complex process through which synoptic magnetograms are derived from line profiles, it is difficult to immediately point out why the two magnetograms differ so drastically.

Vector magnetograms are considered unreliable at high latitudes (\citealp[e.g.][p.~5]{BraPetSin17}; \citealp[p.~2]{LuoJiaWan23}) and in regions where the magnetic field is weak.
The former is related to projection effects, while the latter is due to magnetic effects on the observed lines being overpowered by noise.
The simplest way of obtaining results that do not depend on these regions is to set the magnetic field to zero there.
\emph{Apodization} refers to removing the effect of high-latitude regions by either setting them to zero or using some sort of windowing function.
In figure \ref{bihelical: fig: HMI effect apodization masking}, we show that the peak at low wavenumbers is suppressed when the magnetic field is set to zero either at high latitudes or in weak-field regions.

Many studies \parencite[e.g.][]{BraPetSin17, PipPevLiu19} use magnetograms without any special treatment of the high-latitude regions; one expects such studies to be affected by the instrument-dependence of the strength of the polar fields.
On the other hand, there are also more complicated approaches that involve using an extrapolation scheme to fill in the data at high latitudes \parencite[e.g.][p.~2]{LuoJiaWan23}, which we do not explore here.

\section{Disagreement between helicity spectra from cotemporal HMI and SOLIS magnetograms}
\label{bihelical: section: HMI SOLIS helspec}

\begin{figure*}
	\centering
	\includegraphics{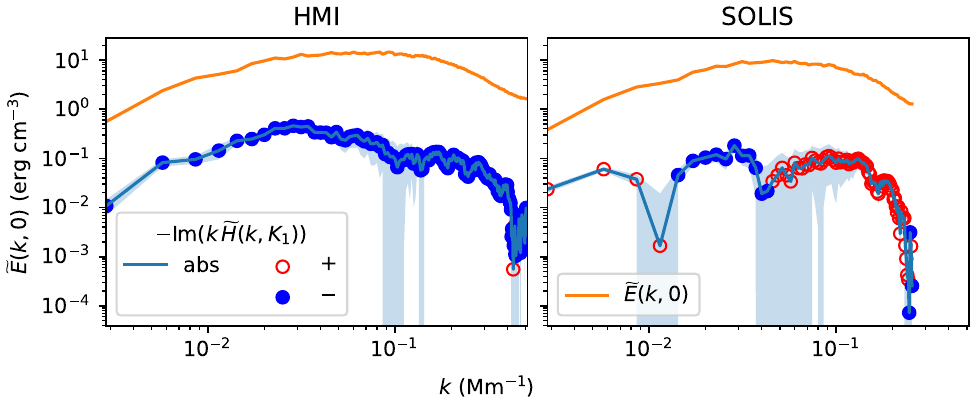}
	\caption{
		Magnetic energy and helicity spectra calculated from apodized ($\abs{\lambda} < \ang{60}$) HMI and SOLIS magnetograms, averaged over Carrington rotations 2142--2151.
		Errors have been indicated by shaded regions of the same colour as the corresponding lines.
		The meanings of the symbols are the same as in figure \ref{B3.bihelical.validation: fig: sim 2}.
		}
	\label{bihelical: fig: compare helspec HMI SOLIS}
\end{figure*}

Despite the issues we noted at large scales, one may still hope that the magnetograms are more reliable at high wavenumbers (which are presumably dominated by active regions, where the strength of the magnetic field allows more accurate inversions).
Figure \ref{bihelical: fig: compare helspec HMI SOLIS} compares energy and helicity spectra from apodized ($\abs{\lambda} < \ang{60}$) HMI and SOLIS magnetograms (Carrington rotations 2142--2151).
As noted by \textcite{SinKapBra18}, we find that the sign of the helicity spectrum is instrument-dependent.
While the interval compared by \textcite[fig.~9]{SinKapBra18} (Carrington rotations 2160--2162) showed disagreement between HMI and SOLIS only at the largest wavenumbers, the particular interval we have compared here shows disagreement at both large and small wavenumbers.

\begin{figure*}
	\centering
	\hfill%
	\includegraphics{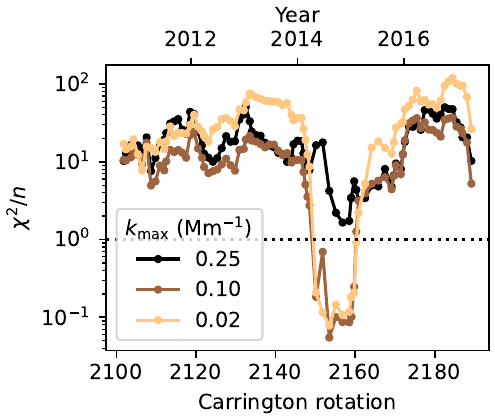}%
	\hfill%
	\includegraphics{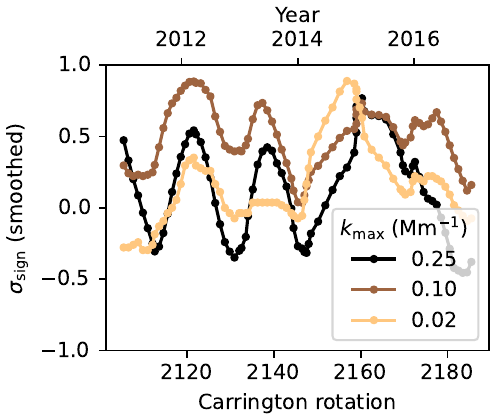}
	\hfill%
	\caption{
		Left:
		reduced $\chi^2$ statistic (equation \ref{bihelical: eq: chi2 defn}) as a function of the Carrington rotation, showing that the differences between spectra from HMI and SOLIS magnetograms are much larger than the estimated errors.
		In each line, only wavenumbers below the indicated cutoff are summed over (the number of such wavenumbers, denoted by $n$, is used to normalize the statistic).
		The cutoff for the black line, \qty{0.25}{Mm^{-1}}, is the Nyquist wavenumber for spectra calculated from the SOLIS synoptic magnetograms.
		Right:
		the $\sigma_\text{sign}$ statistic (equation \ref{bihelical: eQ: sigma_sign_werr defn}), smoothed with a boxcar of width 9 Carrington rotations.
		We see that even the sign of the helicity spectrum is not consistently the same when computed from HMI and SOLIS magnetograms.
		}
	\label{bihelical: fig: chi2 and sigma_sign_werr}
\end{figure*}

To check if this instrument-dependence is consistent with the estimated errors in the helicity spectra, we compute the following statistic:
\begin{equation}
		\chi^2
		\defn
		\sum_k \frac{\left( f_H(k) - f_S(k) \right)^2}{\err_H^2(k) + \err_S^2(k)}
		\label{bihelical: eq: chi2 defn}
\end{equation}
where $f(k) \defn - k \im( \Hspec(k, K_1) ) / \Espec(k,0)$ is the scale-dependent fractional helicity; the subscripts $H$ and $S$ indicate the use of HMI and SOLIS data respectively; and $\err$, the error in $f$, is estimated by propagating the estimated errors in $\Hspec(k, K_1)$ and $\Espec(k,0)$.\footnote{
We follow the usual procedure of adding the relative errors in quadrature.
}
The left panel of figure \ref{bihelical: fig: chi2 and sigma_sign_werr} shows this quantity as a function of the Carrington rotation.
The spectra and errors that determine $\chi^2$ are calculated using a rolling window of width 10 Carrington rotations, and the Carrington rotation attributed to a particular window is the average of the Carrington rotations of all the contributing synoptic magnetograms.
We use synoptic magnetograms for Carrington rotations 2097 through 2195 (this being the interval in which synoptic magnetograms are available from both HMI and SOLIS).
Carrington rotations 2099, 2107, 2127, 2139, 2152, 2153, 2154, 2155, 2163, 2164, 2166, 2167, and 2192 are excluded, since the corresponding SOLIS magnetograms suffer from poor data coverage.

We see that throughout the time period under consideration, the differences between the fractional helicity spectra from HMI and SOLIS synoptic magnetograms are much larger than the errors we have estimated.
If only large wavenumbers ($k < \qty{0.1}{Mm^{-1}}$) are considered, the agreement is similarly bad, except in Carrington rotations 2150--2160 (roughly May 2014 -- February 2015).
It is unclear what is special about this interval of time; no changes in the versioning keywords of the HMI or SOLIS magnetograms seem to coincide with it.

Since the sign of the helicity spectrum is itself of theoretical importance, we also consider another quantity that tells us if at least both instruments predict consistent signs for the helicity spectrum (ignoring any differences in its magnitude):
\begin{equation}
	\sigma_\text{sign}
	\defn
	\meanBr{\sgnerr[\im( \Hspec_H(k, K_1) )] \sgnerr[\im( \Hspec_S(k, K_1) )] }_k
	\label{bihelical: eQ: sigma_sign_werr defn}
\end{equation}
where
\begin{equation}
	\sgnerr(x)
	\defn
	\begin{dcases}
		-1 \,,& \phantom{-\err_x < {}} x < - \err_x
		\\
		\phantom{-{}} 0 \,,& - \err_x < x < \phantom{-{}} \err_x
		\\
		\phantom{-{}} 1 \,,& \phantom{-{}} \err_x < x
	\end{dcases}
\end{equation}
with $\err_x$ being the error in $x$.
The right panel of figure \ref{bihelical: fig: chi2 and sigma_sign_werr} shows that even this quantity (computed using the same magnetograms as above, and further smoothed temporally to bring out broad trends) is not consistently high.
In fact, if the entire range of wavenumbers is considered (i.e.\@ $\kmax = \qty{0.25}{Mm^{-1}}$), this quantity fluctuates about zero, suggesting no correlation on average.

\section{Conclusions}
\label{bihelical: section: conclusions}
While the two-scale method introduced by
\textcite{BraPetSin17} 
may allow one to extract much more information from magnetograms than competing methods, it has certain subtle implementation issues which were not discussed in the original papers.
To aid application of this method, we have discussed these problems and their solutions.

Application of this method to HMI synoptic vector magnetograms reveals that throughout the solar cycle, the magnetic energy spectrum peaks at a wavenumber of roughly \qty{0.1}{Mm^{-1}}.
Near the cycle minimum, a second peak appears at a lower wavenumber, corresponding to Sun's well-known polar fields.
However, somewhat surprisingly, these polar fields are not prominent in cotemporal SOLIS synoptic vector magnetograms.
Previous applications of the two-scale method \citep{BraPetSin17, SinKapBra18} only used magnetograms from the declining phase of solar the solar cycle (where the magnetic energy spectrum has a single peak), and hence this discrepancy was not noticed.

Even after excluding high-latitude regions where the magnetograms are expected to be unreliable, comparison of the helicity spectra computed using the two-scale method reveals that in general, the SOLIS and HMI data differ on both the sign of the magnetic helicity spectrum and the value of the fractional magnetic helicity at all scales.
The helicity spectra at large wavenumbers agree well only in the interval of time between Carrington rotations 2150--2160, which is in the early part of the declining phase of solar cycle 24.
It is unclear if this is just a coincidence.

Part of the disagreement between HMI and SOLIS could, in principle, be attributed to the fact that they observe different spectral lines, and are thus sensitive to magnetic fields at different depths.
However, \citet[section 3.2]{NorGraUlr06} have shown that the magnetic field strengths inferred from these lines agree reasonably well if they are observed simultaneously from the same instrument and analysed in the same way.

Other possible explanations for the disagreement are differences in the way the synoptic magnetograms are assembled, including differences in the observation times of the underlying full-disk magnetograms; and differences in the inversion procedure used to obtain the magnetic field \citep[for a comparison of different inversion procedures, see][fig.~2]{BorLitLag14}.
Directly analysing full-disk magnetograms would allow one to find out which of these is the dominant source of systematic error in the helicity spectrum.
Using full-disk magnetograms would also avoid the loss of information, on two-point correlations, which is inherent to the process of producing synoptic magnetograms (heuristically, the magnetic field present in synoptic magnetograms is already an average over multiple full-disk observations).

In light of our analysis of the two-scale method, the instrument-dependence of other ways of constraining the magnetic helicity should also be critically examined.

\begin{acknowledgments}
	GK thanks Robert Cameron for comments on a preliminary version of this work.
	We are grateful to the referee for helping us improve the manuscript.
	We also thank Alexandra Elbakyan for facilitating access to scientific literature.
	%
	We have used full-disk synoptic vector magnetograms produced by HMI, which is on board the Solar Dynamics Observatory (SDO). SDO is a mission for NASA’s Living with a Star program.
	%
	We have also used SOLIS data obtained by the NSO Integrated
	Synoptic Program (NISP), managed by the National Solar
	Observatory, which is operated by the Association of
	Universities for Research in Astronomy (AURA), Inc., under
	a cooperative agreement with the National Science Foundation.
\end{acknowledgments}

\vspace{5mm}
\facilities{
	SDO (HMI), SOLIS (VSM)
}

\software{
	Astropy \citep{astropy2013, astropy2018, astropy2022},
	DRMS \citep{GloBobCho19},
	Matplotlib \citep{matplotlib2007},
	NumPy \citep{numpy2020},
	Pencil \citep{Pencil2021},
	SciPy \citep{scipy2020}.
}

\appendix

\section{Issues with previous applications of the two-scale method}
\label{bihelical: section: mistakes previous work}

\subsection{A spurious sign flip}
\label{B3.bihelical: section: helspec_two mistake}

We now describe the source of a spurious minus sign in the IDL scripts used by \textcite{SinKapBra18} to analyse SOLIS vector magnetograms.
The specific scripts which we mention by name are available in the git repository associated with our paper, under the folder \lstinline{archaeology}.

Denoting the latitudinal extent by $L_\theta$, we note that in their \lstinline{helspec_two.pro} (lines 15--17), they apply a shift of $L_\theta/2$ along the latitudinal direction to each component of the magnetic field before taking the Fourier transform.
This means that the double correlation $\meanBr{\FT{B}_i(\vec{k}+\vec{K}) \, \FT{B}_j^*(\vec{k})}$, which they calculate, gains an extra multiplicative factor of $e^{i\pi} = -1$ when $\vec{K} =  \left( 2\pi / L_\theta\right) \uvec{\lambda}$.
The shift applied in \lstinline{helspec_two.pro} is unnecessary because the both the SOLIS and HMI vector magnetograms already have the equator at the middle.

Since the minus sign introduced here is cancelled by the minus sign described in section \ref{B3.bihelical.valication: section: traditional}, we believe their results are not affected by this issue.
However, this is something future reproducers of their work need to be aware of.

\subsection{Error estimates}

\Textcite[eq.~11]{SinKapBra18} mention that they estimate errors in a spectrum $P_{k,t}$ as
\begin{equation}
	\sigma^{(P)}_{k} = \sqrt{\meanBr{\left( P_{k,t} - \meanBr{P_{k,t} }_t \right)^2}_t}
	\label{B3.bihelical.arch: eq: singh 2018 error claimed}
\end{equation}
where $t$ is a discrete coordinate meant to represent the Carrington rotation number of the corresponding synoptic map.
However, examination of their IDL routine \lstinline{pavgfew_spec.pro} suggests that they have in fact estimated the error as
\begin{equation}
	\sigma^{(P)}_{k} = \sqrt{\meanBr{\left( \widetilde{P}_{k,t} - \meanBr{\widetilde{P}_{k,t} }_t \right)^2}_t}
	\label{B3.bihelical.arch: eq: singh 2018 error actual}
\end{equation}
where
\begin{equation}
	\widetilde{P}_{k,t} \defn \frac{ P_{k,t-1} + P_{k,t} }{2}
\end{equation}
$\widetilde{P}$ at the beginning of the time slice is defined to be equal to the one at the next time.

Note that equation \ref{B3.bihelical.arch: eq: singh 2018 error claimed} is an estimator of the sample standard deviation, while we are interested in the error in the sample mean.
Equation \ref{B3.bihelical.arch: eq: singh 2018 error actual} appears to be an attempt to estimate the latter and is a variant of the `jackknife' method.

\section{The error due to azimuthal disambiguation}
\label{bihelical: appendix: azimuth error}

\begin{figure*}
	\centering
	\includegraphics{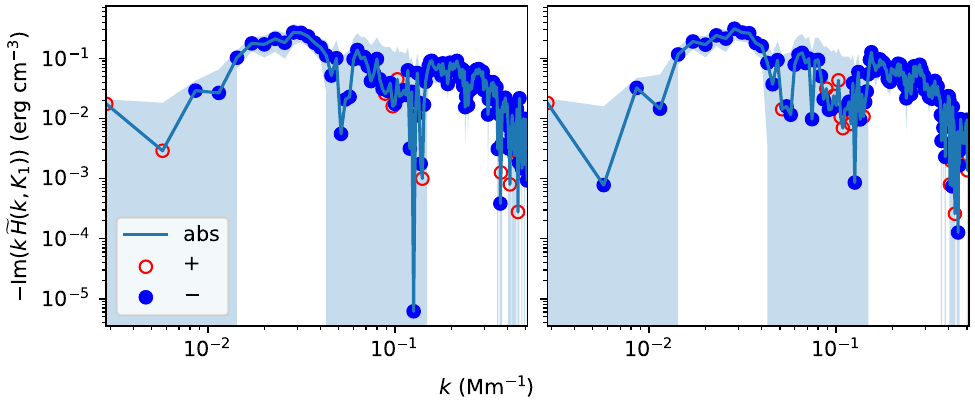}
	\caption{
		Magnetic helicity spectra calculated from different realizations of HMI magnetograms (in each case, the helicity spectra are averaged over Carrington rotations 2143--2152).
		Errors have been indicated by shaded regions of the same colour as the corresponding lines.
		The meanings of the symbols are the same as in figure \ref{B3.bihelical.validation: fig: sim 2}.
		}
	\label{bihelical: fig: helspec error azimuth}
\end{figure*}

Recall that in HMI magnetograms (section \ref{bihelical: section: HMI magnetograms}), the sign of the transverse component of the magnetic field (i.e.\@ the component normal to the line of sight) is chosen randomly in weak-field regions.
One way to estimate the effect of this on the helicity spectra is to construct different realizations of each magnetogram by randomly flipping the sign of the transverse component at each position where the magnitude of the magnetic field is below \qty{150}{G} (the same threshold used to generate the HMI magnetograms).

Since each pixel in a synoptic magnetogram is computed by taking contributions from full-disk magnetograms where the corresponding pixels are close to the central meridian \citep{LiuHoeSun17}, it seems reasonable to assign to each pixel an effective line of sight that points in the direction $\left( \cos\lambda \right) \uvec{r} - \left( \sin\lambda \right) \uvec{\lambda}$.
At each weak-field pixel, we then randomly flip the direction of the component transverse to the corresponding effective line of sight.

Figure \ref{bihelical: fig: helspec error azimuth} compares helicity spectra from two such realizations, showing that the changes in the helicity spectrum are consistent with the error estimated using equation \ref{bihelical: eq: error estimate}.

\bibliography{refs}{}
\bibliographystyle{aasjournal}

\end{document}